\documentclass[prb,twocolumn,aps,superscriptaddress]{revtex4-2}

\usepackage{graphicx}
\usepackage{dcolumn}
\usepackage{bm}
\usepackage{amssymb}
\usepackage{amsmath}
\usepackage{subfigure}
\usepackage{physics}
\usepackage{sublabel}
\usepackage[utf8]{inputenc}
\usepackage{hyperref}
\usepackage[usenames, dvipsnames]{color}
\usepackage[english]{babel}
\usepackage[T1]{fontenc}
\usepackage{bbm}

\usepackage{float}
\hypersetup{colorlinks = true, linkcolor=blue, citecolor=blue, urlcolor=blue}

\begin{document}

\author{Sagnik Banerjee}
\thanks{These authors contributed equally}
\affiliation{Department of Electronics and Telecommunication Engineering, Jadavpur University,  Jadavpur-700032, India
}%

\author{Koustav Jana}
 \thanks{These authors contributed equally}
 \author{Anirban Basak}
 \affiliation{Department of Electrical Engineering, Indian Institute of Technology Bombay, Powai, Mumbai-400076, India
}%
\author{Michael S. Fuhrer}
\affiliation{School of Physics and Astronomy, Monash University, Clayton, Victoria 3800, Australia}
\affiliation{ARC Centre of Excellence in Future Low-Energy Electronics Technologies (FLEET), Monash University, Clayton, Victoria 3800, Australia}
 \author{Dimitrie Culcer}
 \affiliation{School of Physics, University of New South Wales, Sydney 2052, Australia}
\affiliation{ARC Centre of Excellence in Future Low-Energy Electronics Technologies (FLEET), University of New South Wales, Sydney 2052, Australia}
 \author{Bhaskaran Muralidharan}%
 \email{bm@ee.iitb.ac.in}
\affiliation{Department of Electrical Engineering, Indian Institute of Technology Bombay, Powai, Mumbai-400076, India
}%

\title{Robust Subthermionic Topological Transistor Action via Antiferromagnetic Exchange}





\begin{abstract}
The topological quantum field-effect transition in buckled 2D-Xenes can potentially be engineered to enable sub-thermionic transistor operation coupled with dissipationless ON-state conduction. 
Substantive device design strategies to harness this will necessitate delving into 
the physics of the quantum field effect transition between the dissipationless topological phase and the band insulator phase. 
Investigating workable device structures, we uncover fundamental sub-threshold limits posed by the gating mechanism that effectuates such a transition, thereby emphasizing the need for innovations on materials and device structures. Detailing the complex band translation physics related to the quantum spin Hall effect phase transition, it is shown that a gating strategy to beat the thermionic limit can be engineered at the cost of sacrificing the dissipationless ON-state conduction. It is then demonstrated that an out-of-plane antiferromagnetic exchange introduced in the material via proximity coupling can incite transitions between the quantum spin-valley Hall and the spin quantum anomalous Hall phase, which can ultimately ensure the topological robustness of the ON state while surpassing the thermionic limit. Our work thus underlines the operational criteria for building topological transistors using quantum materials that can overcome the Boltzmann’s tyranny while preserving the topological robustness.    

\end{abstract}

\maketitle

\section{Introduction}
A fundamental challenge today in the evolution of field-effect transistors (FETs) is the compulsory power penalty resulting from a fundamental thermionic limit, also known as the Boltzmann's tyranny. This relates to the steepness of the transfer characteristics: the subthreshold swing ($SS$)~\cite{lowpower1,lowpower2,lowpower3}, which is conventionally restricted to 60mV/dec at room temperature. In the context of low-power devices, it is hence paramount to innovate strategies to suppress the $SS$, thereby ensuring a sub-thermionic operation. Several attempts have been made to overcome this limit, popular ones include tunnel FETs~\cite{eeinteractions1,eeinteractions2,eeinteractions3}, impact ionization MOSFETs~\cite{impactionization} and negative capacitance FETs  (NC-FETs)~\cite{negcap1,negcap2,negcap3,negcap4,negcap5,negcap6,negcap7,fuhrer2022proposal}, to name a few.\\
\indent The topological quantum field effect~\cite{nadeem2021overcoming} (TQFE) induced by the Rashba interaction in buckled 2D materials like 2D-Xenes has been recently shown to potentially propel a steeper $SS$ ($SS$<k\textsubscript{B}Tln(10)/q) via a faster-than-linear translation of the topological gap with electric-field modulation. In this context, the topological quantum field-effect transistor (TQFETs)~\cite{topologicalelectronics,ezawa1,ezawa2,topologicaltransistor1,topologicaltransistor2,topologicaltransistor3} should additionally feature the robust dissipationless edge modes hosted in the topological insulator (TI) phase as means toward high ON current and energy-efficient low-power electronics. A great deal of experimental effort has thus already been geared towards the realization of topological transistors~\cite{topologicalexperiments1,topologicalexperiments2,topologicalexperiments3}. Quantum spin Hall (QSH) materials such as group-IV and V-Xenes with buckled 2D honeycomb lattices~\cite{xene1,xene2,xene3,xene4,xene5}, monolayer transition metal dichalcogenides in the 1T' configuration~\cite{TMD-QSH}, HgTe nanoribbons~\cite{HgTenanoribbons} and thin films of 3D topological insulators Bi\textsubscript{2}Se\textsubscript{3}~\cite{bi2se31,bi2se32} as well as Dirac semi-metals like  Na\textsubscript{3}Bi~\cite{Na3Bi} are among the prominent material candidates. \\
\indent  The demonstration of a workable device design uniting the merits of a dissipationless channel along with the possibility of overcoming Boltzmann's tyranny using the TQFE can thus underline the operational criteria for designing TQFETs as a building block for low-power electronics.
In this work, we present such a framework for a holistic analysis of TQFETs and highlight the engineering intricacies involved in harnessing the Rashba spin orbit interaction (SOI) for a steep SS, while simultaneously preserving the topological robustness of the ON state. Building on this, we propose a device structure that utilizes the spin quantum anomalous Hall (SQAH) state induced via an out-of-plane antiferromagnetic (AF) exchange to achieve the desired performance. \\
\begin{figure*}
    \centering
    \includegraphics[scale=0.4]{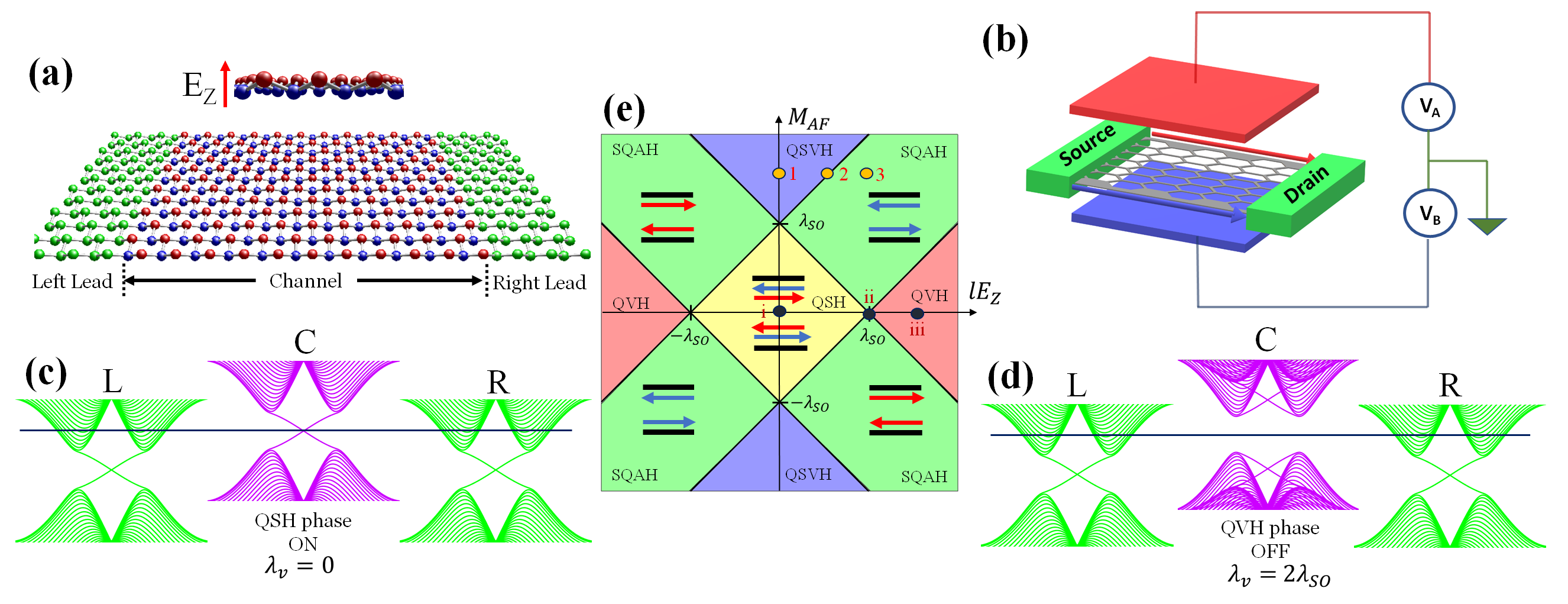}
    \caption{Device structure and phase transitions.
        \textbf{a)} A 2D buckled honeycomb monolayer Xene is used as the channel (C), left (L), and right (R) lead material. The leads are colored green to distinguish them from the channel region, where the two sub-lattices A(B) are represented in red and blue respectively.
    \textbf{b)} Schematic representation of a dual-gated device structure. The top and bottom gates have applied potentials $V_A$ and $V_B$ respectively so that the net potential difference across the channel material is ($V_{A}-V_{B}$). For the symmetric bias arrangement, $V_A = \lambda_v/2$ and $V_B = -\lambda_v/2$.
    \textbf{c, d)} Band structures of the channel and leads for the TQFET, in the ON and the OFF-state respectively. The Fermi level ($E_f$) is represented by a thick solid black line.
    \textbf{e)} Phase diagram of a monolayer Xene nanoribbon with homogeneous perpendicular electric field and antiferromagnetic exchange field~\cite{QSVH_SQAH_2017,QSVH-SQAH_Zhang_2020,QSVH_SQAH_ezawa2013}.
    }
    \label{fig:topo_device_phase}
\end{figure*}
\indent \textcolor{black}{Employing the Keldysh non-equilibrium Green's function (NEGF) formalism~\cite{negf1,negf2,negf3}, we uncover fundamental sub-threshold limits posed by the gating mechanism that effectuates such a transition. By presenting an in-depth analysis of the band translations necessitated by the field effect, we demonstrate that the thermionic limit of the $SS$ in the TQFETs, designed according to conventional principles, is half as steep as that of the conventional FETs i.e., k\textsubscript{B}Tln(10)/q ~\cite{nadeem2021overcoming}.} In an attempt to alleviate this issue, we propose to engineer the gate biasing to modulate one of the bands while restricting the other, ultimately attaining an $SS$ transcending the thermionic limit of 60 mV/dec at room temperature. However, this also introduces dissipative conduction modes from the bulk in the ON state, defeating one of the desired attributes in a TQFET. As a tactical solution, we demonstrate that the introduction of out-of-plane antiferromagnetic exchange interaction, which can be induced via proximity coupling~\cite{proximitycoupling} restores the dissipationless ON state and can effectively reap the merits expected from the Rashba-assisted TQFET. 
\section{Results and Discussion}
 The building block of the TQFET is the buckled hexagonal lattice structure that forms the channel as depicted in Fig.~\ref{fig:topo_device_phase}(a).  
 In the transistor setup depicted in Fig.~\ref{fig:topo_device_phase}(b), an electric field ($E_Z$) applied perpendicular to a buckled channel manifests as a staggered potential between the sub-lattice $A$ and $B$ of the honeycomb unit cell. The dual-gate structure helps in realizing an electric field between the two plates, hence imparting a capacitive action. Such a dual-gate manifestation of a topological transistor also enables a two-fold biasing scheme: a) symmetric biasing, where equal and opposite bias voltages are applied to the two gates, i.e., $V_A=-V_B$, and b) rigid biasing, where the entire voltage is applied to one of the gate plates with the other plate grounded. For the symmetric bias setup the Fermi level is positioned as in Fig~\ref{fig:topo_device_phase}(c) and Fig~\ref{fig:topo_device_phase}(d) to ensure topological ON state conduction.
 \\
\indent Previous works have been centered around the electric-field driven transition, i.e., the transition involving the QSH phase, which effectively navigates the horizontal axis of the phase diagram in Fig.~\ref{fig:topo_device_phase}(e). By modulating the strength of the AF exchange interaction $M_{AF}$, the phase transition could also track the vertical axis. In the former, the phase transition is between the conducting QSH phase and the insulating quantum valley Hall (QVH) phase. In the latter, phase transitions are between an insulating quantum spin valley Hall (QSVH) phase and a conducting spin quantum anomalous Hall (SQAH) phase~\cite{QSVH_SQAH_2017,QSVH-SQAH_Zhang_2020,QSVH_SQAH_ezawa2013}. 
Unlike the former case, here the AF interaction breaks the time reversal symmetry (TRS), resulting in spin-polarized conducting modes without a chiral counter-propagating partner.

\indent To analytically investigate the subthreshold physics of topological transistors we adopt the low energy effective four-band Bloch Hamiltonian $H_{\eta}$ in the vicinity of Dirac points $K(K')$ as given by~(\ref{Heq_Dirac}). This Dirac Hamiltonian has been derived from the tight-binding Hamiltonian model for a 2D buckled honeycomb lattice as elaborated in the Supplementary Information (refer to \eqref{Heq}). 
\begin{multline} \label{Heq_Dirac}
H_{\eta} = \hbar v_f(\eta k_x \tau_x + k_y \tau_y)\sigma_0 + \eta             \lambda_{SO} \tau_z \sigma_z \\
       + \lambda_v\left(E_Z\right)\tau_z\sigma_0 + \frac{\lambda_R\left(E_Z\right)}{2}(\eta\tau_x\sigma_y - \tau_y\sigma_x),
\end{multline}
where $\eta=+(-)$ is the valley index denoting $K(K')$ and, $\sigma$ and $\tau$ are the spin and pseudo-spin Pauli matrices respectively. Here, $\hbar$ is the reduced Planck's constant and $v_f$ denotes the Fermi velocity, given by the expression $v_{f} = 3ta_{o}/2$ where $t$ is the hopping parameter and $a_{o}$ is the lattice constant. The quantities $\lambda_{SO}$, $\lambda_{v}$ and $\lambda_{R}$ denote the strengths of the intrinsic spin-orbit coupling, staggered sublattice potential and Rashba spin-mixing interaction respectively. Additional details about the same can be found in the Supplementary Information. 
\\
\begin{figure}
    \centering
    \includegraphics[scale=0.4]{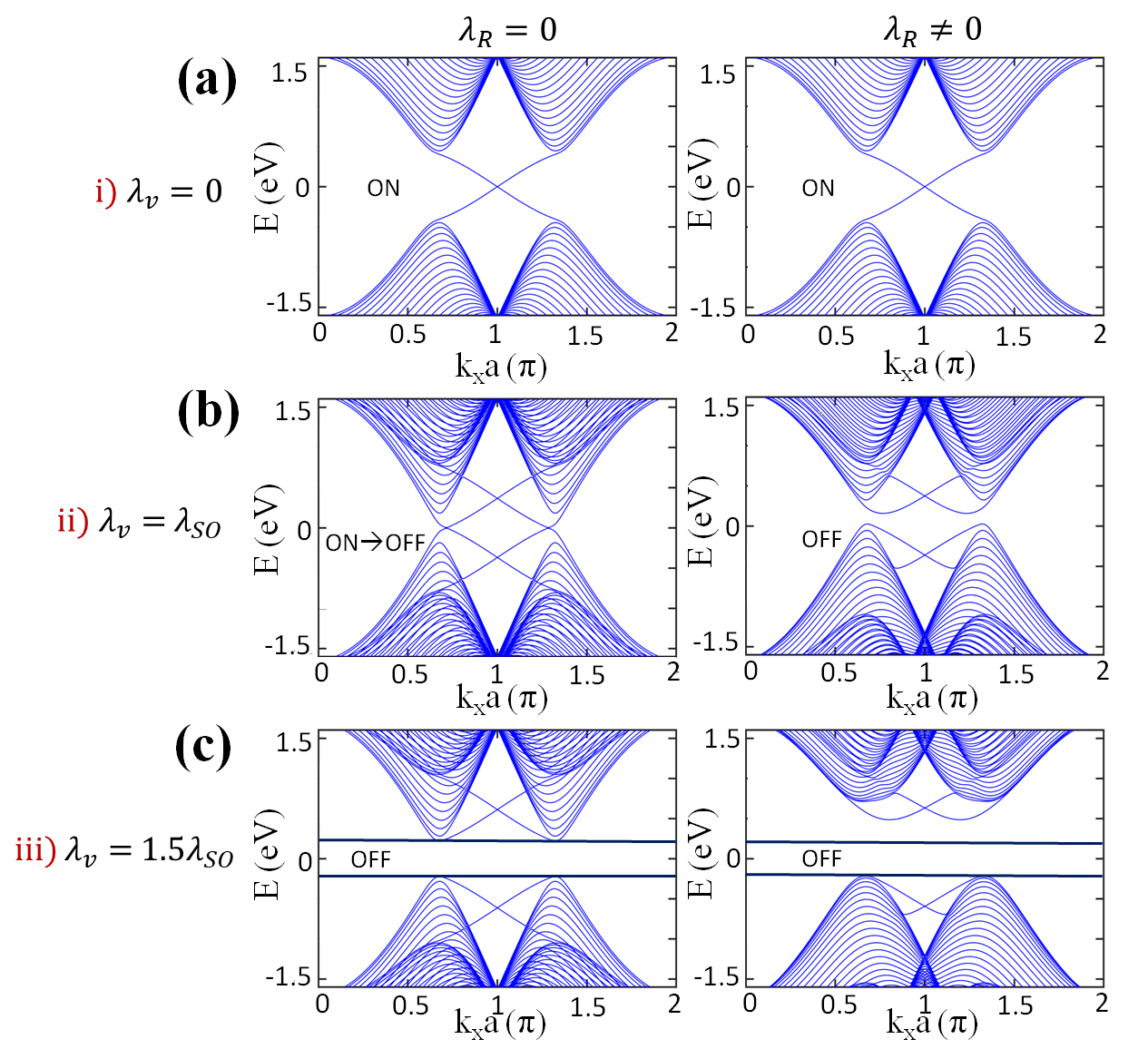}
    \caption{Band structures of the channel for different $\lambda_v$.
        \textbf{a)} represents the QSH phase characterized by topologically protected edge states. Here, we have assumed $\lambda_v=0$ and considered two cases of $\lambda_R=0$ and $\lambda_R \neq 0$. 
        \textbf{b)} depicts the transition from QSH to QVH phase for the $\lambda_R=0$ case. For the  $\lambda_R \neq 0$ case, the system has already entered the band insulator phase. Here, we have considered $\lambda_v=\lambda_{SO}$, where $\lambda_{SO}$ is the intrinsic spin-orbit coupling strength.
        \textbf{c)} demonstrates a faster retreat of the CB in the $\lambda_R \neq 0$ case compared to the $\lambda_R = 0$ case. The VB in both cases remains almost pinned. Here, $\lambda_v=1.5\lambda_{SO}$.
    }
    \label{fig:topo_qsh_qvh_band}
\end{figure}
\indent In the context of an electric field induced topological phase transition between the QVH and the QSH phases \cite{Basak_2021}, it is important to analyze the Chern number \cite{jana2021robust} ($\mathcal{C}$) and the nature of the edge states. For $\lambda_v<\lambda_{SO}$, the QSH phase is characterized by a zero total Chern number $\mathcal{C}$ and a non-zero spin Chern number $2\mathcal{C}_s = +2$ resulting in helical edge states as illustrated in Fig.~\ref{fig:topo_qsh_qvh_band}(a). A phase transition at $\lambda_v=\lambda_{SO}$, as in Fig.~\ref{fig:topo_qsh_qvh_band}(b), results in the QVH phase for $\lambda_v>\lambda_{SO}$. This band insulator phase is characterized by both zero $\mathcal{C}$ and $2\mathcal{C}_s$, leading to a trivial gap as shown in Fig.~\ref{fig:topo_qsh_qvh_band}(c). The faster closing and reopening of gap for the $\lambda_{R} \neq 0$ case when compared to $\lambda_{R} = 0$ case, as shown in Fig.~\ref{fig:topo_qsh_qvh_band}, nicely illustrates the topological quantum field effect (TQFE) switching. \\ 
\indent The total current with components $I_c$ and $I_v$ due to the electrons in the conduction band (CB) and holes in the valence band (VB) is given as
\begin{equation} \label{currents}
\begin{split}
& I_c = I_{co}\exp-\frac{q\left(E_c-E_f\right)}{k_BT}\\
& I_v = I_{vo}\exp\frac{q\left(E_v-E_f\right)}{k_BT},
\end{split}
\end{equation}
where $I_{co(vo)}$ is the CB (VB) current maximum, $E_{c(v)}$ represents the CB (VB) minimum (maximum), and $E_f$ denotes the equilibrium Fermi energy level. Here, $k_BT$ is the thermal energy at temperature $T$.\\
\indent The $SS$ of a transistor during the ON-OFF state transition is hence defined as
\begin{equation} \label{subthrshold_swing}
SS = \left|\frac{d(log_{10}I)}{dV_G}\right|^{-1},
\end{equation}
where $I$ ($=I_c+I_v$) represents the current in response to an applied gate voltage $V_G$. Substituting $I$ from (\ref{currents}) in (\ref{subthrshold_swing}), we get,
\begin{equation} \label{SS}
SS = \frac{k_BT}{q}ln(10)\frac{I}{\left|I_v\left(\frac{dE_v}{dV_G}\right)-I_c\left(\frac{dE_c}{dV_G}\right)\right|},
\end{equation}
where 2k\textsubscript{B}Tln(10)/q is the thermionic limit at temperature $T$, which reduces to 60 mV/decade at room temperature. We then define the reduced $S^{*}$, which is the reduced $SS$ \cite{nadeem2021overcoming} as,
\begin{equation} \label{S*}
S^{*} = \frac{I}{\left|I_v\left(\frac{dE_v}{dV_G}\right)-I_c\left(\frac{dE_c}{dV_G}\right)\right|}.
\end{equation}
For conventional MOSFETs, it has hitherto been assumed that $S^{*}\geq1$. Moreover, it is anticipated in \cite{nadeem2021overcoming} that the introduction of Rashba interactions can achieve $S^{*}\leq1$ to overcome the Boltzmann's limit. However, we now demonstrate that this limit ($S^*$) for a TQFT is, unfavorably, twice as much.\\
\begin{figure}
    \centering
    \includegraphics[scale=0.4]{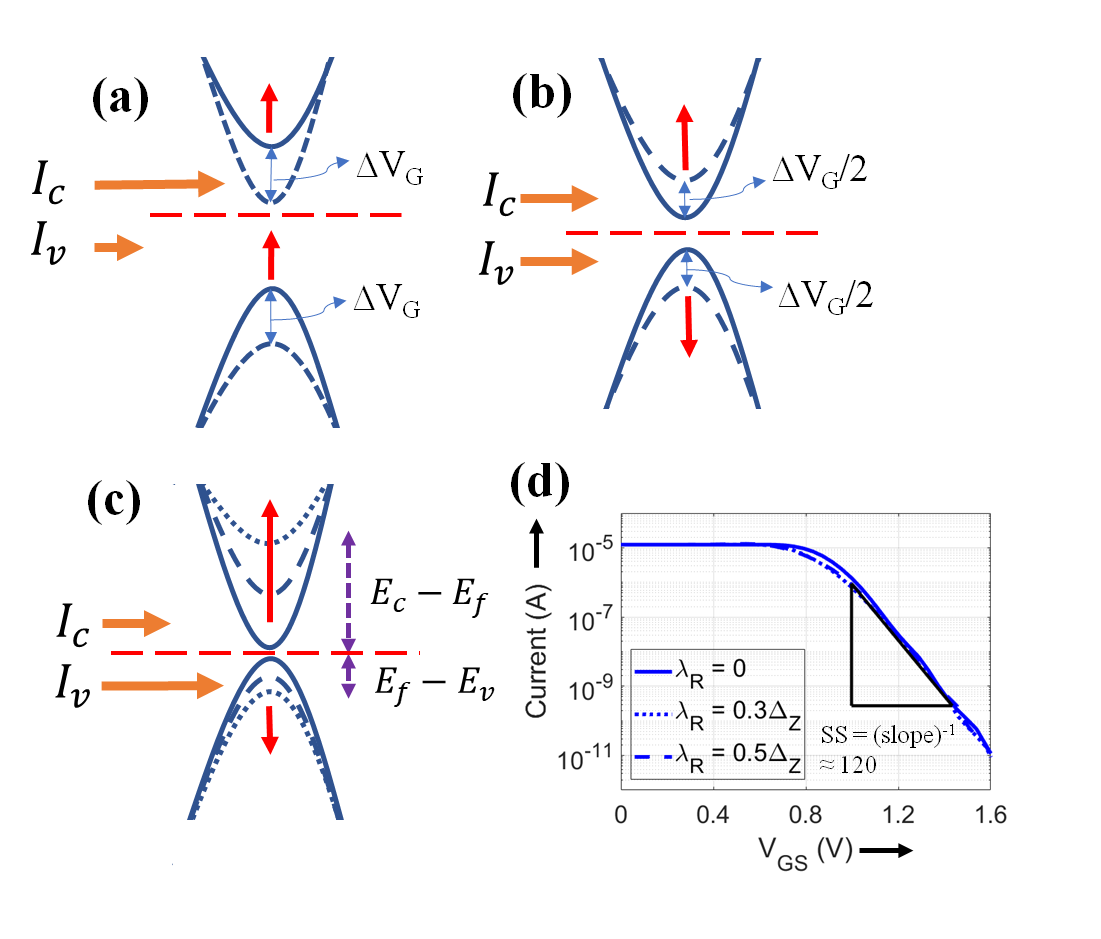}
    \caption{Band movement schematics and I-V characteristics for QSH-QVH transition based FET.
        \textbf{a)} represents the band movement in an ordinary n-MOSFET for an applied bias potential of $\Delta V_G$. Here, $I_C>I_V$
        \textbf{b)} represents the band movement in the OFF state of a topological transistor with $\lambda_R=0$. Here, $I_C=I_V$.
        \textbf{c)} represents the band movement in the OFF state of a topological transistor with $\lambda_R \neq 0$. Here, $I_C<I_V$.
        \textbf{d)} is the I-V characteristics of a topological transistor based upon phase transition from ON ($V_{GS}<0.8$V) to OFF ($V_{GS}>0.9$V) phase, for different Rashba strengths of $\lambda_R=0$, $\lambda_R=0.3\Delta_Z$ and $\lambda_R=0.5\Delta_Z$. Here, $\Delta_Z=\lambda_v$. For all the three cases, the subthreshold-swing ($SS$) remains confined to 120 mV/decade of current.
    }
    \label{fig:topo_SS_Analytical}
\end{figure}
\indent Figure \ref{fig:topo_SS_Analytical}(a) depicts the n-MOS operation of a conventional transistor. For an applied gate bias $\Delta V_G$, both the CB and the VB move equally by an amount proportional to $\Delta V_G$. For an n-MOS device, this upward ascent of the bands results in $I_c>>I_v$. In this case, both $(dE_c/dV_G)$ and $(dE_v/dV_G)$ are unity since both bands translate by an equal amount ($\propto \Delta V_G$).\\
\indent For a TQFET with $\lambda_R=0$, however, the translation of bands in response to $\Delta V_G$ is non-trivial. As depicted in Fig. \ref{fig:topo_SS_Analytical}(b), both the bands move away from the Fermi level by an amount proportional to $\Delta V_G/2$ in the OFF state. So, $(dE_c/dV_G)=1/2$ and $(dE_v/dV_G)=-1/2$ for a TQFET with $\lambda_R=0$. Moreover, due to symmetry in the band translation, the current components $I_c \approx I_v \approx I/2$. This results in $S^* \approx 2$ and hence the $SS$ in a standard TQFET will be restricted to 2k\textsubscript{B}Tln(10)/q instead of k\textsubscript{B}Tln(10)/q. At room temperature, this translates to the thermionic limit being restricted to 120 mV/decade instead of 60 mV/decade in the conventional case.\\
\indent With the introduction of the Rashba SOI term, the band translation during the OFF state (QVH phase) is asymmetric. From the Dirac Hamiltonian elaborated in \eqref{Heq_Dirac}, we calculate $E_c$ and $E_v$ for $\lambda_v>\lambda_{SO}>0$, i.e., in the OFF state as follows:
\begin{equation} \label{eigenvalues}
\begin{split}
& E_v = \lambda_{SO} - \lambda_v \\
& E_c = -\lambda_{SO} + \sqrt{{\lambda_v^2} + \lambda_R^2},
\end{split}
\end{equation}
where, $\lambda_v$ and $\lambda_R$ can be represented as $\alpha_V E_Z$ and $\alpha_R E_Z$ respectively since they are both linearly proportional to the gate electric field $E_Z$. Taking derivatives with respect to $V_G$ we get, $(dE_v/dV_G)=-1/2$ and $(dE_c/dV_G)=0.5(1+\alpha_R^2/\alpha_V^2)^{1/2}$.
Note that, putting $\alpha_R=0$ we get same results as estimated previously for the $\lambda_R=0$ case. Also, we notice that in the QVH phase, the CB translates faster with increasing gate bias than the VB due to the effect of the Rashba term $\alpha_R$. However, since $\left|E_v\right|<\left|E_c\right|$, we expect $I_v$ to contribute more to the current than $I_c$. Thus, from \eqref{S*} we expect $(dE_v/dV_G)$ to contribute more to the $SS$. Thus, the improved $SS$ due to  $(dE_c/dV_G)$ is undone by the relatively smaller contribution from $I_c$. This mechanism has been represented schematically in Fig. \ref{fig:topo_SS_Analytical}(c). 
\\
\indent In Fig. \ref{fig:topo_SS_Analytical}(d), we show the evaluated subthreshold characteristics using the NEGF approach ~\cite{negf1,negf2,negf3,LNEB}. We set $T=300$K and $\lambda_{SO}=0.41$eV with a hopping parameter $t=1.6$eV, consistent with first-principles calculations ~\cite{nadeem2021overcoming}. We assume a finite-sized nanoribbon as the channel material, with a device length $N_L=45$ and width $N_W=20$. The NEGF simulations reconcile with the analytical conclusions drawn. Despite the introduction of Rashba interactions, the $SS$ is limited to 2k\textsubscript{B}Tln(10)/q or 120 mV/decade at $T=300$K. 
\begin{figure}
    \centering
    \includegraphics[scale=0.4]{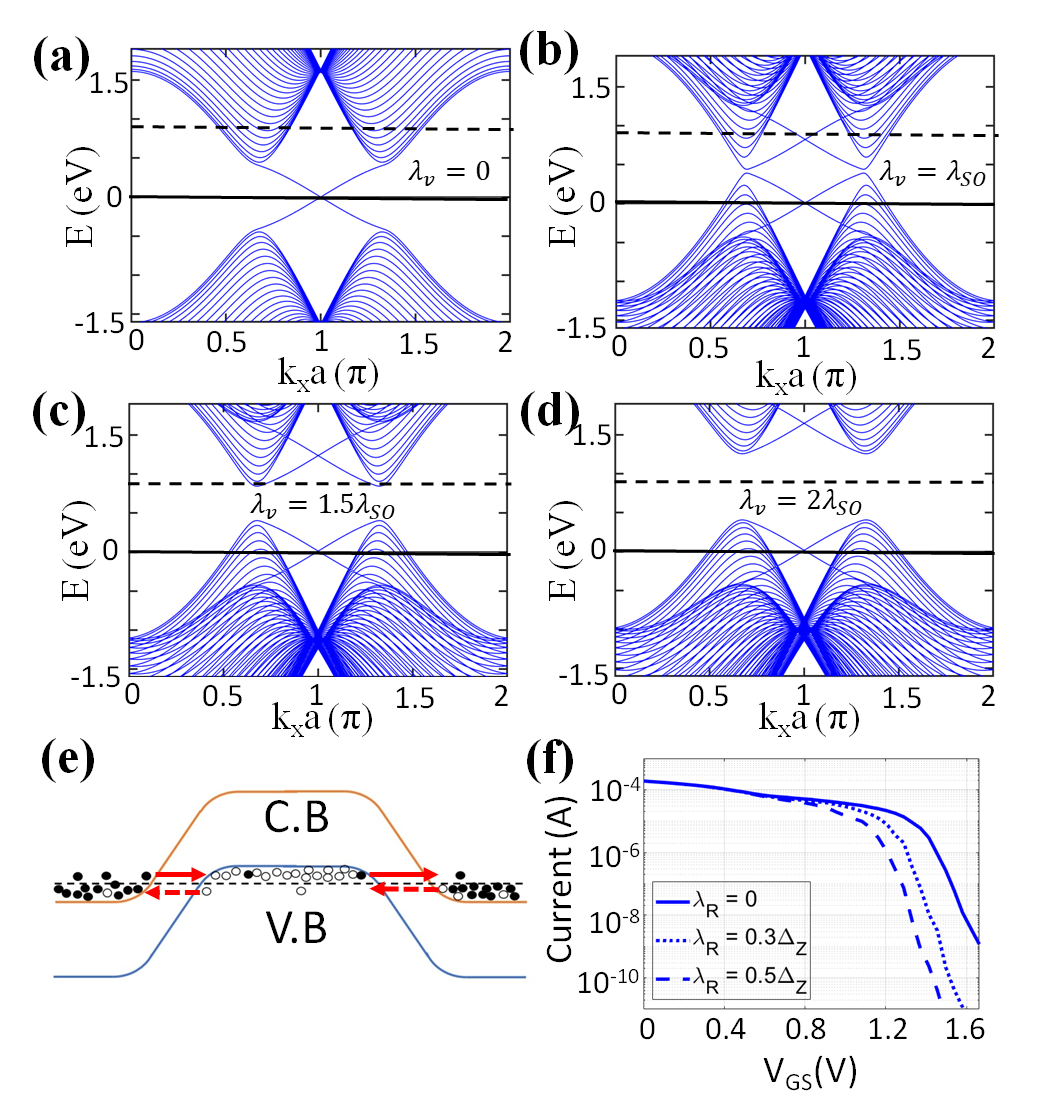}
    \caption{Rigid biasing approach wherein $V_A=V_G$ and $V_B=0$. The band structures for varying $\lambda_v$ along with the corresponding Fermi levels (elaborated in text) are depicted in \textbf{(a-d)}. \textbf{(e)} describes the band-to-band tunneling possibility when the channel Fermi level lies inside the VB, leading to significant amount of OFF current. \textbf{(f)} is the I-V characteristics for a rigidly biased TQFET with suitably positioned Fermi level for different values of $\lambda_R$. The corresponding $SS$ values for $\lambda_R = 0$, $0.3\Delta_z$, and $0.5\Delta_z$ are 60mV/dec, 52mV/dec, and 49mV/dec respectively 
    }
    \label{fig:topo_contact_eng}
\end{figure}
With the symmetric gate biasing, as seen in previous sections, the channel Fermi level is pinned in the middle of the bandgap of the channel. As a result, in the QVH phase (OFF) the channel conducts the smallest possible current.
Moreover, using symmetric gate biasing, the $SS$ cannot be improved significantly below 120mV/dec even in the presence of Rashba SOI. Hence, an alternate approach is required which relies upon a smart positioning of the channel Fermi level to create an imbalance in the rate with which the two bands move away from the channel Fermi level, as a function of the applied gate voltage. \\ 
\indent The easiest way to do this would be via an asymmetric gate biasing where $V_A$ and $V_B$ corresponding to Fig.~\ref{fig:topo_device_phase}(b) are set to $V_G$ and zero respectively. \textcolor{black}{This particular gating scheme has an added advantage that it is compatible with the modern-day transistor architectures because of it requiring only a single gate voltage as opposed to the dual gate voltage requirement of symmetric gate biasing.} Now positioning the Fermi level at $E=0$ (solid line in Fig.~\ref{fig:topo_contact_eng}(a-d)) leads to an immanent problem as described below. At $\lambda_v = 0$, as shown in Fig.~\ref{fig:topo_contact_eng}(a), the Fermi level lies right at the middle of the bandgap. Because of rigid biasing in the QSH phase, only the VB moves with varying gate voltage and after a certain point, the Fermi level slips inside the VB. Beyond the critical field, the Fermi level remains pinned inside the VB. As a result, the TQFET does not turn OFF in the QVH phase, because of the VB bulk states participating in the conduction. Despite the Fermi level in the leads being aligned to the CB, the current conduction is still facilitated by band-to-band tunneling as depicted in Fig.~\ref{fig:topo_contact_eng}(e). Hence, one obtains a large OFF current which is undesirable.  \\
\indent \textcolor{black}{The problem highlighted above arises because of the Fermi level lying inside the VB in the QVH phase resulting in a deteriorated OFF state. One obvious way to alleviate this issue is to align the Fermi level inside the bandgap for the QVH phase. However, in this scenario, the Fermi level moves inside the CB in the QSH phase (ON state). Thus, apart from the dissipationless edge modes, even the dissipative bulk states participate in conduction.
This strategy restores the full advantage of the Rashba SOI-enabled TQFE in reducing the SS. However, conduction no longer occurs solely through dissipationless edge states. This may be disadvantageous, though the details of the ON state conduction will depend on material and device parameters, and in principle dissipative bulk states can also contribute to enhanced ON current. Thus, by allowing dissipative bulk state conduction, it is possible to achieve subthermionic performance with negligible OFF current, by aligning the Fermi level as shown by the dashed lines in Fig.~\ref{fig:topo_contact_eng}(a-d).}\\
\indent The obtained I-V characteristics for the above case are as shown in Fig.~\ref{fig:topo_contact_eng}(f). 
The ON current is significantly higher (> $10^4$) when compared to that of the symmetric biasing case in Fig.~\ref{fig:topo_SS_Analytical}(d), owing to the fact that the current-carrying states in the ON state are now the bulk modes of the QSH phase. Because of rigid gate biasing, as illustrated by our numerical simulations, the $SS$ calculated is around 60mV/dec, the thermionic limit at 300K. The inclusion of Rashba SOI further enhances the subthreshold performance with $SS$ values of 52mV/dec and 50mV/dec for the cases of $\lambda_R = 0.3\Delta_z$ and $0.5\Delta_z$ respectively, thus exemplifying subthermionic performance enabled by the introduction of Rashba SOI for a rigidly biased TQFET with a suitably positioned Fermi level. 
\\
\begin{figure}
    \centering
    \includegraphics[scale=0.4]{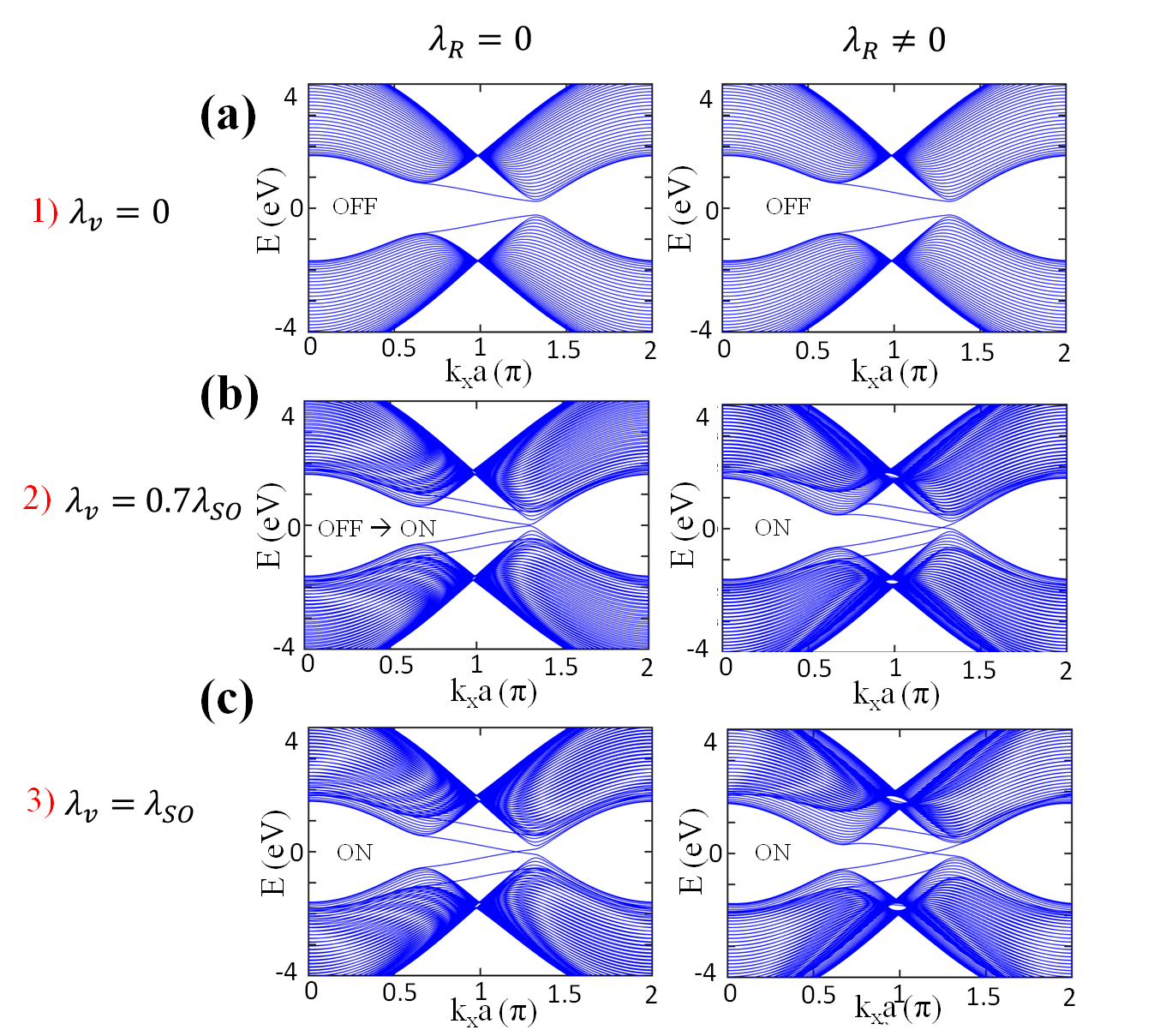}
    \caption{Band structures of the channel for different $\lambda_v$ highlights the underlying switching principle under the influence of an antiferromagnetic term.
        \textbf{a)} represents the quantum spin valley Hall (QSVH) phase characterized by an insulating gap for the two valleys K(K'). Here, we have assumed $\lambda_v=0$ and considered two cases of $\lambda_R=0$ and $\lambda_R \neq 0$. 
        \textbf{b)} depicts the QSVH-SQAH transition for the $\lambda_R=0$ case. For the  $\lambda_R \neq 0$ case, topologically protected edge states are prominent. Here, we have considered $\lambda_v=0.5\lambda_{SO}$, $\lambda_{SO}$ being the intrinsic spin-orbit coupling strength.
        \textbf{c)} demonstrates the onset of ON phase for both the $\lambda_R=0$ and $\lambda_R\neq0$ case. Here, we have chosen $\lambda_v=0.8\lambda_{SO}$.
    }
    \label{fig:topo_qsvh_sqah_band}
\end{figure}
\indent \textcolor{black}{While the above strategy is a robust route to a subthermionic transistor, we would also like to understand whether it is possible to retain both the subthermionic SS and the dissipationless edge transport in the ON state.
Dissipationless edge transport should provide significant advantages for certain device geometries and hence, is an indispensable feature to retain.
We thus explore ideas beyond the QSH-QVH transition and look into new topological phases. 
As suggested by \eqref{SS}, one way to achieve $S^{*} \leq 1$ is by ensuring that at least one of the quantities $(dE_c/dV_G)$ and $(dE_v/dV_G)$ exceeds unity. If, say, the CB satisfies this, then for subthermionic performance the CB current should be the major contributor of the total current. In other words, we can attain subthermionic performance by ensuring $(dE_c/dV_G) \geq 1$ and $I_c >> I_v$.} \\
\begin{figure}
    \centering
    \includegraphics[scale=0.4]{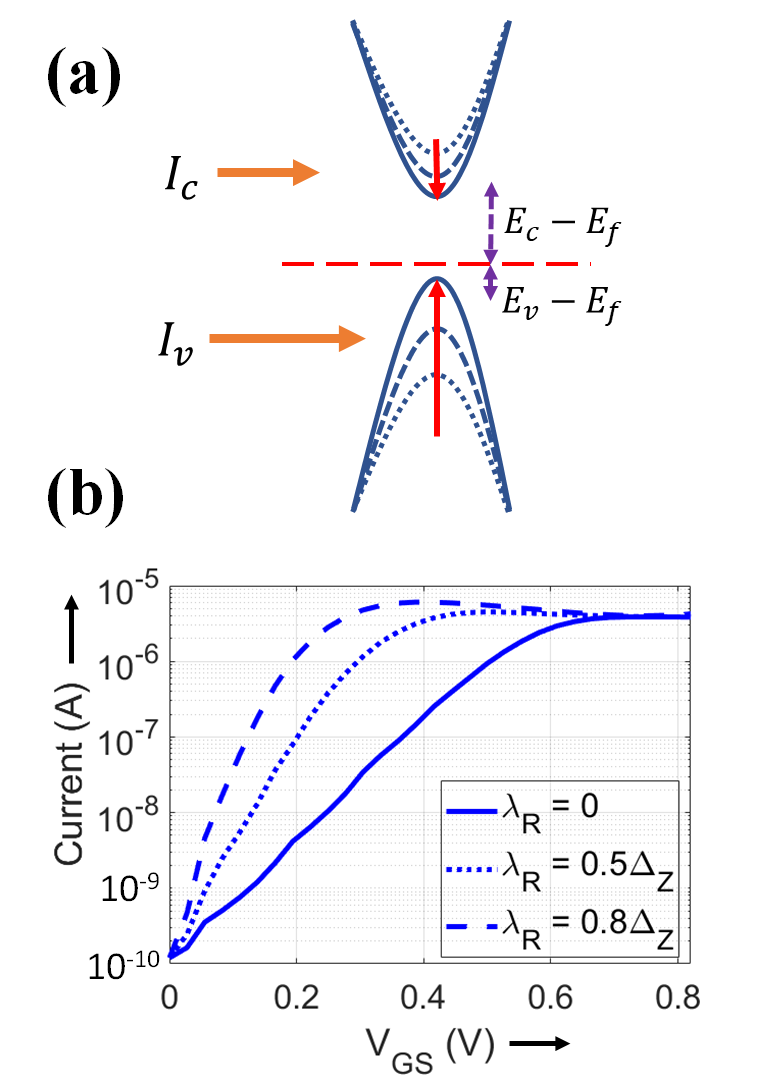}
    \caption{
        \textbf{a)} The band translation schematics in the OFF state of a TQFET is represented which includes AF exchange interactions, when $I_C < I_V$. The Fermi level is closer to the VB than the CB. \textbf{b)} The I-V characteristics based on phase transitions in the subthreshold region for different Rashba interactions, for $\Delta_Z= \lambda_v$. We notice a progressive improvement in the $SS$ as $\lambda_R$ increases. 
    }
    \label{fig:topo_AFM_SS_Analytical}
\end{figure}
\indent In the QSH-QVH transition-based TQFET with symmetric biasing, evident from \eqref{eigenvalues}, it is possible to achieve  $(dE_c/dV_G) \geq 1$ in the presence of a large enough Rashba SOI. However as a result of this, in the QVH phase, the CB moves away from the Fermi level at a greater rate than the VB and we get $I_v >> I_c$, as illustrated in Fig.~\ref{fig:topo_SS_Analytical}(c), and thus there is no discernible improvement in $S^{*}$. This happens because in the OFF state the bandgap opens as we increase the gate voltage. For the Rashba influenced band to contribute more to the total current, the bandgap should close as the applied field increases. In such a scenario, the Rashba influenced band would be closer to the Fermi level throughout the subthreshold regime (OFF state). This will eventually lead to a subthermionic performance.
\\ 
\indent The above scenario can be realized by the addition of AF exchange interaction, which can be incorporated via proximity coupling~\cite{proximitycoupling} to a 2D topological insulator implying the addition of the term $M_{AF}\tau_{z}\sigma_{z}$ to the low-energy Hamiltonian in \eqref{Heq_Dirac}. As represented by the points (1),(2) and (3) in Fig.~\ref{fig:topo_device_phase}(e), for $M_{AF} > \lambda_{SO}$, one can realize the QSVH phase and the SQAH phases by varying the perpendicular electric field~\cite{QSVH_SQAH_2017,QSVH-SQAH_Zhang_2020,QSVH_SQAH_ezawa2013}. For $\lambda_v < M_{AF}-\lambda_{SO}$, we obtain the QSVH phase with a Chern number $\mathcal{C} = 0$ and no gapless edge states. On the other hand, $\lambda_v > M_{AF}-\lambda_{SO}$ opens the SQAH phase with a Chern number $\mathcal{C} = 1$, possessing spin-polarized chiral edge states.
\textcolor{black}{The perfect spin polarization of the edge currents may also provide connections to spintronics applications.}
Here, the QSVH phase represents the OFF state, i.e., the transition is from OFF to ON as the gate voltage increases, contrary to the scenario in the QSH-QVH transition. \\
\indent Now, as depicted in Fig.~\ref{fig:topo_AFM_SS_Analytical}(a), the bandgap closes in the subthreshold regime on increasing gate voltage and the Rashba influenced VB moves closer to the Fermi level, such that $I_v >> I_c$. In this case, if $\lambda_R$ is large enough for $dE_v/dV_G$ to be greater than unity, it is possible to achieve $S^{*}<1$.  Analytically, based on the low-energy Dirac Hamiltonian in \eqref{Heq_Dirac}, and in presence of Rashba SOI, we have the following in the QSVH phase:
\begin{equation} \label{eigenvalues_AFM}
\begin{split}
& E_v = \lambda_{SO} - M_{AF} + \sqrt{{\lambda_v^2} + \lambda_R^2} \\
& E_c = M_{AF} -\lambda_{SO} -\lambda_v.
\end{split}
\end{equation}
Clearly, on increasing $E_z$, the VB approaches the Fermi level at a greater rate for a non-zero $\lambda_R$, and the VB current is the major contributor. 
\\
\indent These analytical findings are well supported by the simulated results in Fig.~\ref{fig:topo_AFM_SS_Analytical}(b). Here the following parameters are considered: $\lambda_{SO} = 0.41eV$, $M_{AF} = 1.7\lambda_{SO}$. We consider three values of $\lambda_R$ for comparison: $\lambda_R = 0$, $0.5\Delta_z$ and $0.8\Delta_z$, where $\Delta_z=\lambda_v$. The $\lambda_R = 0$ case, similar to the previous QSH-QVH case, gives an $SS$ of around 120mV/dec, because of symmetric band translation under symmetric biasing. However, upon increasing $\lambda_R$, we notice a significant improvement in the $SS$, with the topological transistor achieving $SS <$ 60mV/dec for $\lambda_R = 0.8\Delta_z$. Thus the addition of an out-of-plane AF exchange interaction to a 2D-TQFET under symmetric biasing can attain subthermionic characteristics. One thing to note is that the current remains constant in the ON state for an appreciable range of gate voltage, suggesting that the dissipationless edge modes are responsible for the ON state conduction. Adding to this, a TQFET hosting chiral QAH edge modes is expected to be more resilient to back-scattering than the one having helical QSH edge modes. This is because QSH requires the time-reversal symmetry (TRS) to be preserved for maintaining its robustness. However, this is not the case with the QAH phase which is robust even to back-scattering by magnetic disorder, because of the absence of any time-reversal partner for the chiral edge modes. 
\section{Conclusion}
Our analysis into the physics of the quantum field effect transition unraveled that the fundamental subthreshold performance of the QSH-QVH transition is at best half as steep as that of the conventional field-effect transistor and that the mere introduction of Rashba interaction does not render any additional steepness. We proposed tactical upgrades to alleviate these drawbacks to actually steer toward the desired subthermionic performance. We first demonstrated that a modified gating scheme could drive the topological transition and successfully overcome the thermionic limit while sacrificing the dissipationless nature of the ON state. We then proposed to exploit the topological transition between the QSVH and the SQAH phase via the introduction of out-of-plane AF exchange interaction, thereby ensuring the topological robustness of the ON state while surpassing the thermionic limit. Our work thus underlines the operational criteria for building topological transistors using quantum materials that can overcome the Boltzmann’s tyranny while preserving the topological robustness.

\section{Methods}
All the transport calculations are based on the NEGF formalism~\cite{negf1,negf2,negf3,LNEB}, within the tight binding framework of the model Hamiltonian \cite{ezawa1,ezawa2,Basak_2021,jana2021robust} described in the Supplementary Information. 
To obtain the desired I-V characteristics, current calculations are performed based on the Landauer transmission formula, evaluated from the device retarded Green's function
\begin{equation}
    \label{Itot}
    I = \frac{e^2}{h}\int_{-\infty}^{\infty} T(E) (f(E,\mu_{L},T) - f(E,\mu_{R},T))dE,
\end{equation}
where $f(E,\mu,T) = (1+exp(\frac{E-\mu}{k_{B}T}))^{-1}$ is the Fermi-Dirac distribution at Fermi energy $\mu$ and temperature $T$. The calculation of the transmission coefficient $T(E)$ using NEGF formalism has been discussed in the Supplementary Information. 

\section*{Acknowledgements}
The author BM acknowledges the Visvesvaraya Ph.D Scheme of the Ministry of Electronics and Information Technology (MEITY), Government of India, implemented by Digital India Corporation (formerly Media Lab Asia). The author BM also acknowledges the support by the Science and Engineering Research Board (SERB), Government of India, Grant No. Grant No. STR/2019/000030, and the Ministry of Human Resource Development (MHRD), Government of India, Grant No. STARS/APR2019/NS/226/FS under the STARS scheme. MSF and DC acknowledge support from the ARC Centre of Excellence in Future Low-Energy Electronics Technologies (CE170100039).


\onecolumngrid
\section*{Supplementary Information}
\subsection{Device Hamiltonian}
\renewcommand{\theequation}{S\arabic{equation}}
\renewcommand{\thefigure}{S\arabic{figure}}
\setcounter{equation}{0}
\setcounter{figure}{0}
 For numerical calculations pertinent to our proposed topological transistor, we consider the typical tight-binding Hamiltonian model for a 2D buckled honeycomb lattice~\cite{tightbinding1,tightbinding2}, as shown in (\ref{Heq}) in the second quantized notation: 
\begin{multline} \label{Heq} 
\hat{H}  = -t\sum_{\langle i,j \rangle\alpha}^{} c_{i\alpha}^{\dagger}c_{j\alpha} +
            i \frac{\lambda_{SO}}{3\sqrt{3}}\sum_{\langle \langle i,j \rangle \rangle \alpha \beta}^{} \nu_{ij}c_{i\alpha}^{\dagger}s_{\alpha \beta}^{z}c_{j\beta}
            +\lambda_v\sum_{i\alpha}^{}c_{i\alpha}^{\dagger}\mu_{i}c_{i\alpha}+i\lambda_R\sum_{ij\langle \alpha \beta \rangle}^{}c_{i\alpha}^{\dagger}(s_{\alpha \beta}\times\hat{d}_{ij})_{z}c_{j\beta},
\end{multline}
where $c_{i\alpha}^{(\dagger)}$ represents the electronic annihilation (creation) operator on site $i$ with a spin $\alpha = \uparrow (\downarrow)$, and
$\langle i,j \rangle$ and $\langle \langle i,j \rangle \rangle$ characterize the nearest neighbour and the next-nearest neighbour hopping respectively. The spin indices are represented with corresponding values $+1/-1$ respectively. The first term in \eqref{Heq} represents the nearest-neighbor hopping term with a hopping strength $t$. The second term represents the intrinsic spin-orbit (SO) coupling with strength $\lambda_{SO}$, where $\nu_{ij} = +1 (-1)$ for anti-clockwise (clockwise) next-nearest neighbour hopping with respect to the positive $z$-axis.
The third term denotes the staggered sub-lattice potential of strength $\lambda_v$, with $\mu_{i} = +1(-1)$, where $i$ denotes the sub-lattice A(B). The fourth term represents the nearest neighbour Rashba spin-mixing interaction, with $s_{\alpha\beta}$ denoting the corresponding matrix elements indicating spin-polarization $\alpha$, $\beta$ at lattice sites $i$, $j$ respectively, and \textbf{$\hat d_{ij}$} is the distance vector between lattice sites $i$ and $j$. In the low-energy limit, the essential physics governing the device operation can be captured using the four-band Bloch Hamiltonian around the Dirac points $K(K')$ as in equation (1) of the main article.  \\
\subsection{Calculation of transmission coefficient}
The I-V calculations have been done using equation (8) of the main article, which involves the term $T(E)$ i.e. the coherent transmission coefficient at a given energy $E$. To calculate $T(E)$, we employ the NEGF formalism~\cite{negf1,negf2,negf3,LNEB} based on the tight-binding framework described in \eqref{Heq}, where $T(E)$ is evaluated using the Green's function as:
\begin{equation}
\label{T} 
    T(E) = \mathbf{Tr}[\Gamma_{L}(E)G^{R}(E)\Gamma_{R}(E)G^{A}(E)]
\end{equation}
\begin{equation}
    \label{gam} 
    [\Gamma_{L,R}(E)] = i[\Sigma_{L, R}(E)-\Sigma_{L, R}(E)^{\dag}]
\end{equation}
\begin{equation}
    \label{gr} 
    [G^{R}(E)] = [(E+i\eta)I - H - \Sigma_{L}(E) - \Sigma_{R}(E)]^{-1}
\end{equation}
 where $\mathbf{Tr}$ represents the trace operation, $[\Gamma_{L(R)}(E)]$ is the broadening matrix corresponding to the lead $L(R)$, and $[G^{R}(E)]$ and $[G^{A}(E)]$ are the matrix representations of the retarded and advanced Green's functions respectively. All quantities in the above equations can be obtained from the Hamiltonian defined in \eqref{Heq} and the self-energy matrices $[\Sigma_{L, R}]$, which are calculated recursively based on the formalism prescribed in~\cite{Basak_2021,LNEB}. 

\twocolumngrid
\bibliography{topo_paper_ref}

\end{document}